\documentstyle[12pt]{article}

\expandafter\ifx\csname mathrm\endcsname\relax\def\mathrm#1{{\rm #1}}\fi

\makeatletter
\if@twoside \oddsidemargin -17pt \evensidemargin 00pt
\else \oddsidemargin 00pt \evensidemargin 00pt
\fi
\makeatother

\def\beq{\begin{equation}}
\def\eeq{\end{equation}}
\def\beqar{\begin{eqnarray}}
\def\eeqar{\end{eqnarray}}
\def\barr#1{\begin{array}{#1}}
\def\earr{\end{array}}
\def\bfi{\begin{figure}}
\def\efi{\end{figure}}
\def\btab{\begin{table}}
\def\etab{\end{table}}
\def\bce{\begin{center}}
\def\ece{\end{center}}
\def\nn{\nonumber}

\def\text{\textstyle}
\def\arraystretch{1.2}

\def\al{\alpha}

\def\ga{\gamma}
\def\de{\delta}
\def\eps{\varepsilon}

\def\si{\sigma}

\def\Ga{\Gamma}
\def\De{\Delta}

\def\refeq#1{\mbox{(\ref{#1})}}
\def\reffi#1{\mbox{Fig.~\ref{#1}}}
\def\reffis#1{\mbox{Figs.~\ref{#1}}}
\def\refta#1{\mbox{Tab.~\ref{#1}}}

\def\refse#1{\mbox{Sect.~\ref{#1}}}

\def\citere#1{\mbox{Ref.~\cite{#1}}}
\def\citeres#1{\mbox{Refs.~\cite{#1}}}

\def\solid{\raise.9mm\hbox{\protect\rule{1.1cm}{.2mm}}}
\def\dash{\raise.9mm\hbox{\protect\rule{2mm}{.2mm}}\hspace*{1mm}}

\newcommand{\GeV}{\unskip\,\mathrm{GeV}}
\newcommand{\MeV}{\unskip\,\mathrm{MeV}}
\newcommand{\TeV}{\unskip\,\mathrm{TeV}}

\newcommand{\ord}{{\cal O}}

\def\mathswitchr#1{\relax\ifmmode{\mathrm{#1}}\else$\mathrm{#1}$\fi}

\newcommand{\PW}{\mathswitchr W}
\newcommand{\PZ}{\mathswitchr Z}

\newcommand{\PH}{\mathswitchr H}

\newcommand{\Pl}{\mathswitchr l}

\newcommand{\Pt}{\mathswitchr t}

\newcommand{\PWpm}{\mathswitchr {W^\pm}}
\newcommand{\PWO}{\mathswitchr {W^0}}

\newcommand{\Gl}{\Ga_{\mathrm l}}

\def\mathswitch#1{\relax\ifmmode#1\else$#1$\fi}

\newcommand{\MWpm}{\mathswitch {M_\PWpm}}
\newcommand{\MWO}{\mathswitch {M_\PWO}}

\newcommand{\MZ}{\mathswitch {M_\PZ}}
\newcommand{\MH}{\mathswitch {M_\PH}}

\newcommand{\Mt}{\mathswitch {m_\Pt}}

\renewcommand{\ss}{\scriptscriptstyle}
\newcommand{\sw}{\mathswitch {s_{\ss\PW}}}

\newcommand{\swbar}{\mathswitch {\bar s_{\ss\PW}}}

\newcommand{\GF}{\mathswitch {G_\mu}}

\newcommand{\alpz}{\alpha(\MZ^2)}
\newcommand{\alps}{\alpha_{\mathrm s}}

\newcommand{\bos}{{\mathrm{bos}}}
\newcommand{\fer}{{\mathrm{ferm}}}

\newcommand{\SC}{{\mathrm{SC}}}
\newcommand{\IB}{{\mathrm{IB}}}

\newcommand{\LEP}{{\mathrm{LEP}}}

\def\draftdate{\relax}
\def\mda{\relax}
\def\mua{\relax}
\def\mla{\relax}
\def\draft{
\def\thtystars{******************************}
\def\sixtystars{\thtystars\thtystars}
\typeout{}
\typeout{\sixtystars**}
\typeout{* Draft mode!
         For final version remove \protect\draft\space in source file *}
\typeout{\sixtystars**}
\typeout{}
\def\draftdate{\today}
\def\mua{\marginpar[\boldmath\hfil$\uparrow$]%
                   {\boldmath$\uparrow$\hfil}%
                    \typeout{marginpar: $\uparrow$}\ignorespaces}
\def\mda{\marginpar[\boldmath\hfil$\downarrow$]%
                   {\boldmath$\downarrow$\hfil}%
                    \typeout{marginpar: $\downarrow$}\ignorespaces}
\def\mla{\marginpar[\boldmath\hfil$\rightarrow$]%
                   {\boldmath$\leftarrow $\hfil}%
                    \typeout{marginpar: $\leftrightarrow$}\ignorespaces}
\def\Mua{\marginpar[\boldmath\hfil$\Uparrow$]%
                   {\boldmath$\Uparrow$\hfil}%
                    \typeout{marginpar: $\Uparrow$}\ignorespaces}
\def\Mda{\marginpar[\boldmath\hfil$\Downarrow$]%
                   {\boldmath$\Downarrow$\hfil}%
                    \typeout{marginpar: $\Downarrow$}\ignorespaces}
\def\Mla{\marginpar[\boldmath\hfil$\Rightarrow$]%
                   {\boldmath$\Leftarrow $\hfil}%
                    \typeout{marginpar: $\Leftrightarrow$}\ignorespaces}
\overfullrule 5pt
\oddsidemargin -15mm
\marginparwidth 29mm
}

\makeatletter

\def\eqnarray{\stepcounter{equation}\let\@currentlabel=\theequation
\global\@eqnswtrue
\global\@eqcnt\z@\tabskip\@centering\let\\=\@eqncr
$$\halign to \displaywidth\bgroup\hskip\@centering
  $\displaystyle\tabskip\z@{##}$\@eqnsel&\global\@eqcnt\@ne
  \hskip 2\arraycolsep \hfil${##}$\hfil
  &\global\@eqcnt\tw@ \hskip 2\arraycolsep $\displaystyle\tabskip\z@{##}$\hfil
   \tabskip\@centering&\llap{##}\tabskip\z@\cr}
\def\appendix{\par
 \setcounter{section}{0} \setcounter{subsection}{0}
 \def\thesection{\Alph{section}}}

\makeatother

\begin{document}
\thispagestyle{empty}
\def\thefootnote{\fnsymbol{footnote}}
\setcounter{footnote}{1}
\null
\hfill BI-TP 95/34 \\
\null
\hfill hep-ph/9511281
\vskip .8cm
\begin{center}
{\Large \bf Which Bosonic Loop Corrections are Tested in
Electroweak Precision Measurements?}%
\footnote{Presented by G.~Weiglein at
{\it The Second German-Polish Symposium,
New Ideas in the Theory of Fundamental Interactions}, Zakopane,
September 1995.}%
\footnote{Partially supported by the EC-network contract CHRX-CT94-0579.}
\vskip 3em
{\large S.\ Dittmaier, D.\ Schildknecht and G.\ Weiglein%
\footnote{Supported by the Bundesministerium f\"ur Bildung und Forschung,
Bonn, Germany.}
}
\vskip .5em
{\it Fakult\"at f\"ur Physik, Universit\"at Bielefeld,
D-33615 Bielefeld, Germany}
\vskip 2em
\end{center} \par
\vskip 1.2cm
\vfil

{\bf Abstract} \par
The nature of the electroweak bosonic loop corrections to which current
precision experiments are sensitive is explored. The set of effective
parameters $\De x$, $\De y$, and $\eps$,
which quantify SU(2) violation in an effective Lagrangian,
is shown to be particularly useful for this purpose. The standard
bosonic corrections are sizable only in
the parameter $\De y$, while $\De x$ and $\eps$ are sufficiently well
approximated by the pure fermion-loop prediction. By analyzing the
contributions to $\De y$ it is shown that the bosonic loop corrections
resolved by the present precision data are induced by the change in
energy scale between the low-energy process muon decay and the energy
scale of the LEP1 observables.
If the (theoretical value of the) leptonic width of the W~boson is
used as input parameter instead of the Fermi constant $\GF$,
no further bosonic loop corrections are necessary for compatibility
between theory and experiment.
\par
\vskip 1cm
\noindent October 1995 \par
\null
\setcounter{page}{0}
\clearpage
\def\thefootnote{\arabic{footnote}}
\setcounter{footnote}{0}

\section{Introduction}
\label{intro}
By comparing the results of precision experiments with the theoretical
predictions of the electroweak Standard Model (SM)
in various approximations,
it is possible to test the loop corrections of
this model, i.e.\ to test the SM as a quantized
field theory. Considering the present experimental accuracy, the
question then naturally arises to which radiative corrections the data
are in fact sensitive. While the pure fermion-loop predictions of the
SM involve only couplings that have already been studied in
low-energy experiments (except for
the couplings of the top quark), the
full one-loop predictions of the SM involve also the non-abelian gauge
structure and the Higgs sector, for which much less direct
experimental information is available.

Genuine precision tests of the
electroweak theory therefore require an experimental accuracy that
allows to distinguish between the pure fermion-loop and the full
one-loop predictions of the theory~\cite{GounSchi}.
This accuracy was first reached in 1994, as shown in an
analysis~\cite{zph1,zph2}
incorporating as observables the W-boson mass $\PWpm$ and
the leptonic Z-peak observables $\Ga_\Pl$,
i.e.\ the leptonic Z-boson decay width, and $\bar\sw^2$,
i.e.\ the leptonic effective weak mixing angle,
which are not influenced by the
discrepancies noted in certain hadronic decay modes of the $\PZ$~boson.
Indeed, by systematically
discriminating between fermion-loop
(vacuum-polarization) corrections to the $\ga$, $\PZ$ and $\PWpm$
propagators and the full one-loop results, it was found
that contributions beyond fermion loops are required for consistency
with the experimental results on these observables.
While the pure fermion-loop predictions were shown to
be incompatible with the data,
the complete one-loop prediction of the
SM provides a consistent description of the
experimental results. This implies that the data
have become sensitive to bosonic radiative corrections and thus provide
quantitative tests of the
non-abelian gauge structure of the standard
electroweak theory. The experimental evidence for bosonic loop
corrections was also explored for the single observable $\bar\sw^2$ in
\citere{ga94} and for $\MWpm$ in \citere{hi95}. The evidence for
radiative corrections beyond the $\alpz$-Born approximation, which
takes into account fermion-loop corrections to the photon propagator
only, was explored in \citere{okun}.

The analysis in \citeres{zph1,zph2} is based on an effective
Lagrangian~\cite{zph0} for electroweak interactions that incorporates
possible sources of SU(2) violation in the leptonic sector via three effective
parameters, $\De x$, $\De y$, and $\eps$. They parametrize
SU(2) breaking in the vector-boson masses, in the couplings of the
vector bosons to charged leptons and in the mixing among  the neutral
vector bosons. In the analysis based on this effective Lagrangian the
parameters $\De x$, $\De y$, and $\eps$ are directly related to
observables and are thus manifestly gauge-independent quantities. Their
theoretical predictions incorporate the full SM radiative corrections.
This set of parameters is related
by linear combinations to the parameters $\eps_i (i = 1, 2, 3)$
of \citere{alta}, which were introduced by isolating the
leading terms of the top-quark mass dependence.
Apart from emerging naturally from
symmetry breaking in an underlying effective Lagrangian, the parameters
$\De x$, $\De y$, and $\eps$ are particularly convenient for investigating
the relevance of radiative corrections beyond fermion loops. The
evaluation of the parameters in the SM shows that the bosonic loop
corrections required for consistency with the data are completely
contained in only one of the parameters, namely $\De y$, while $\De x$
and $\eps$ may consistently be approximated by
the pure fermion-loop predictions.
While $\De y$ is at present
the only parameter
in which standard bosonic contributions are significant, it
is totally insensitive to the Higgs
sector of the SM; it does not even show a logarithmic dependence for
a heavy Higgs-boson mass.

In \citere{zph3} the bosonic contributions to $\De y$ have further been
investigated. It has been shown that the bosonic corrections to which
current precision experiments are sensitive can be traced back to a
scale-change effect related to the use of the low-energy parameter
$\GF$, which is measured in muon decay,
for the analysis of the LEP observables.
This fact has explicitly been demonstrated by inserting the SM theoretical
value of the leptonic W-boson width as input instead of $\GF$.

In the present article the aforementioned results are surveyed. The
investigations are based on the most recent data presented at the
1995 Summer Conferences~\cite{data8/95}.
The paper is organized as follows:
In \refse{sec:params} the concept of analyzing the data in terms of the
effective parameters $\De x$, $\De y$, and $\eps$ is briefly sketched.
The fermion loop predictions for these parameters are compared to the
full SM predictions and to the experimental values of these
parameters.
In \refse{sec:dely} it is shown that the bosonic contribution to $\De y$
is strongly dominated by the correction induced by the change in
energy scale from muon decay to the LEP observables.
Supplementing the pure fermion-loop predictions with the
bosonic scale-change contributions to $\De y$ leads to a consistent
description of the observables $\Ga_\Pl$, $\bar\sw^2$, and $\MWpm$.
In \refse{sec:gawscheme} the significance of the bosonic
corrections is discussed
in a scheme where the leptonic W-boson width is taken from
theory and used as an input parameter instead of $\GF$.
Final conclusions are drawn in \refse{conc}.

\section{\boldmath{Data analysis in terms of the effective parameters
$\De x$, $\De y$, and $\eps$}}
\label{sec:params}

The effective Lagrangian introduced for the analysis of LEP1
observables in \citeres{zph1,zph2,zph0} quantifies SU(2)-breaking
effects in the leptonic sector by the parameters
$x = (1 + \De x)$, $y = (1 + \De y)$, and $\eps$. It contains the SM
tree-level form of the
vector-boson fermion interactions
in the limit $\De x = \De y = \eps = 0$.

In the charged-current Lagrangian the $\PWpm$ boson
is coupled to the weak isospin current $j^{\pm}_\mu$
via the coupling $g_\PWpm$,
\beq
{\cal L}_{\mathrm C}=-{1\over 2} W^{+\mu \nu} W^-_{\mu \nu} -
\frac{g_\PWpm}{\sqrt 2}\left( j^+_\mu W^{+\mu} + h.c.\right)
+ \MWpm^2 W^+_\mu W^{-\mu}.
\label{lc}
\eeq
In the transition to the neutral-current sector SU(2) symmetry is
broken in the coupling of the $(W^{\pm},W^0)$ triplet
by introducing the parameter $y$,
\beq
g_{\PWpm}^2 = y g_{\PW^0}^2 = (1 + \De y) g_{\PW^0}^2,
\label{y}
\eeq
and in the mass
terms via the parameter $x$,
\beq
\MWpm^2 = x\MWO^2 = (1+\De x) \MWO^2.
\label{mwpm}
\eeq
SU(2) violation in $\ga W^0$ mixing is furthermore quantified by the
parameter $\eps$ according to
\beq
{\cal L}_{\mathrm mix} = -\frac{1}{2} \frac{e(\MZ^2)}{g_\PWO}
(1-\eps) A_{\mu \nu} W^{0,\mu \nu},
\label{lm}
\eeq
where $e^2(\MZ^2) \equiv 4 \pi \alpz$
denotes the electromagnetic coupling at the
\PZ-boson mass.
In \citeres{zph1,zph2, zph0} the charged-current coupling $g_\PWpm$
was defined with respect to muon decay, i.e.\ at a low-energy scale, as
\beq
\label{eq:gw0}
g^2_\PWpm \equiv g^2_\PWpm(0) \equiv 4\sqrt{2}\GF\MWpm^2,
\eeq
while in the neutral sector $g_{\PW^0}^2 \equiv g_{\PW^0}^2(\MZ^2)$
corresponds to the coupling at the \PZ-boson scale.

After diagonalization the neutral-current Lagrangian
can be written as
\beqar
{\cal L}_{\mathrm N} & = &
-{1\over 4} Z_{\mu \nu} Z^{\mu \nu} +
\frac{\MWpm^2}{2x^{\prime}
\left(1-\bar\sw^2(1-\eps^{\prime})\right)}
Z_\mu Z^\mu   \nn\\[.3em]
&& -\frac{g_\PWpm}{\sqrt{y^{\prime}
\left(1-\bar\sw^2 (1-\eps^{\prime})\right)}}
\left[j^{3}_\mu - \bar\sw^2 j_{{\mathrm em},\mu}\right] Z^\mu,
\label{lnc2}
\eeqar
where the shorthands
\beq
x^{\prime} = x + 2 s^2_0 \de, \qquad
y^{\prime} = y - 2 s^2_0 \de, \qquad
\eps^{\prime}  = \eps - \de,
\label{xyeprime}
\eeq
have been used and the small quantity $\de$
($\de \sim 10^{-4}$ in the SM)
describes parity violation in the photonic coupling at the
Z-boson mass scale (see \citere{zph2}).
The effective weak mixing angle $\bar\sw^2$, which empirically is
determined by the charged lepton asymmetry at the \PZ-boson resonance,
is given as
\beq
\bar\sw^2 = \frac{e^2(\MZ^2)}{g_\PWpm^2(0)}y^{\prime}(1-\eps^{\prime}),
\label{swbar}
\eeq
and $s^2_0$ in (\ref{xyeprime}) is defined via
\beq
s^2_0(1-s^2_0) = s_0^2c_0^2 = {{\pi \alpz}\over{\sqrt 2 G_\mu \MZ^2}}.
\label{s02}
\eeq

Using the effective Lagrangian ${\cal L} = {\cal L}_{\mathrm C} +
{\cal L}_{\mathrm N}$ to express the weak mixing angle $\bar\sw^2$,
the \PWpm\ mass $\MWpm$, and the leptonic
Z-boson width $\Gl$ in terms
of $\De x$, $\De y$, and $\eps$, and linearizing in these parameters
yields
\beqar
\bar\sw^2 &=& s_0^2\left[1-\frac{1}{c_0^2-s_0^2}\varepsilon
-\frac{c_0^2}{c_0^2-s_0^2}(\Delta x-\Delta y)
+(c_0^2-s_0^2)\de
\right], \nn\\
\frac{\MWpm}{\MZ} &=& c_0\left[1+\frac{s_0^2}{c_0^2-s_0^2}\varepsilon
+\frac{c_0^2}{2(c_0^2-s_0^2)}\Delta x
-\frac{s_0^2}{2(c_0^2-s_0^2)}\Delta y
\right], \nn\\
\Gl &=& \Gl^{(0)}\left[1+
\frac{8s_0^2}{1+(1-4s_0^2)^2}\left\{
\frac{1-4s_0^2}{c_0^2-s_0^2}\varepsilon
+\frac{c_0^2-s_0^2-4s_0^4}{4s_0^2(c_0^2-s_0^2)}
(\Delta x-\Delta y)
+2s_0^2\de
\right\}\right], \hspace*{1.9em}
\label{obsleplin}
\eeqar
with
\beq
\Gl^{(0)} = \frac{\alpz\MZ}{48s_0^2c_0^2}
\left[1+(1-4s_0^2)^2\right]\left(1+\frac{3\alpha}{4\pi}\right).
\label{obsleplow}
\eeq

As experimental
input for our analysis we use the most recent experimental
data~\cite{data8/95},
\beqar
\MZ &=& 91.1884 \pm 0.0022 \GeV, \nn\\
\Gl &=& 83.93 \pm 0.14 \MeV,\nn\\
\swbar^2 (\LEP) &=& 0.23186 \pm 0.00034, \nn\\
\frac{\MWpm}{\MZ} ({\mathrm UA2+CDF}) &=& 0.8802\pm0.0018, \nn\\
\alps(\MZ^2) &=& 0.123\pm0.006.
\label{scdata}
\eeqar
We restrict the analysis to the LEP value of $\swbar^2$. Using
instead the combined LEP+SLD value, $\swbar^2=0.23143\pm0.00028$
\cite{data8/95}, does not significantly affect our results.
The data in \refeq{scdata} are supplemented by the Fermi constant
\beq
\GF=1.16639 (2) \cdot 10^{-5} \GeV^{-2},
\label{eq:GF}
\eeq
and the electromagnetic coupling at the $\PZ$-boson resonance,
\beq
\alpz^{-1} = 128.89 \pm 0.09,
\label{alpz1}
\eeq
which was taken from the recent updates~\cite{bu95} of the evaluation of
the hadronic vacuum polarization.

Using these input data and solving \refeq{obsleplow} for $\De x$, $\De y$,
and $\eps$ with a corresponding error analysis yields
as experimental values for the parameters
\beq
\begin{array}[b]{rcrcrcrl}
\De x^{\mathrm exp}     &= (& 10.1 &\pm& 4.2 &\pm& 0.2 &)
\times 10^{-3},\\
\De y^{\mathrm exp}     &= (&  5.4 &\pm& 4.3 &\pm& 0.2 &)
\times 10^{-3},\\
 \eps^{\mathrm exp}      &= (& -5.3 &\pm& 1.6 &\mp& 0.5&)
\times 10^{-3}.
\earr
\label{expxyeps}
\eeq
The first errors indicated in \refeq{expxyeps}
are due to the statistical and
systematic errors in the experimental data, and the second errors give
the deviations caused by the replacement
$\alpz^{-1}\to\alpz^{-1}\pm\de\alpz^{-1}$
according to \refeq{alpz1}.

\begin{figure}
\begin{center}
\begin{picture}(16,16.0)
\put( 6.0,14.8){Fig.\ref{paramfig}a}
\put(14.0,14.8){Fig.\ref{paramfig}b}
\put( 6.0, 6.4){Fig.\ref{paramfig}c}
\put(-2.7, -6.0){\includegraphics{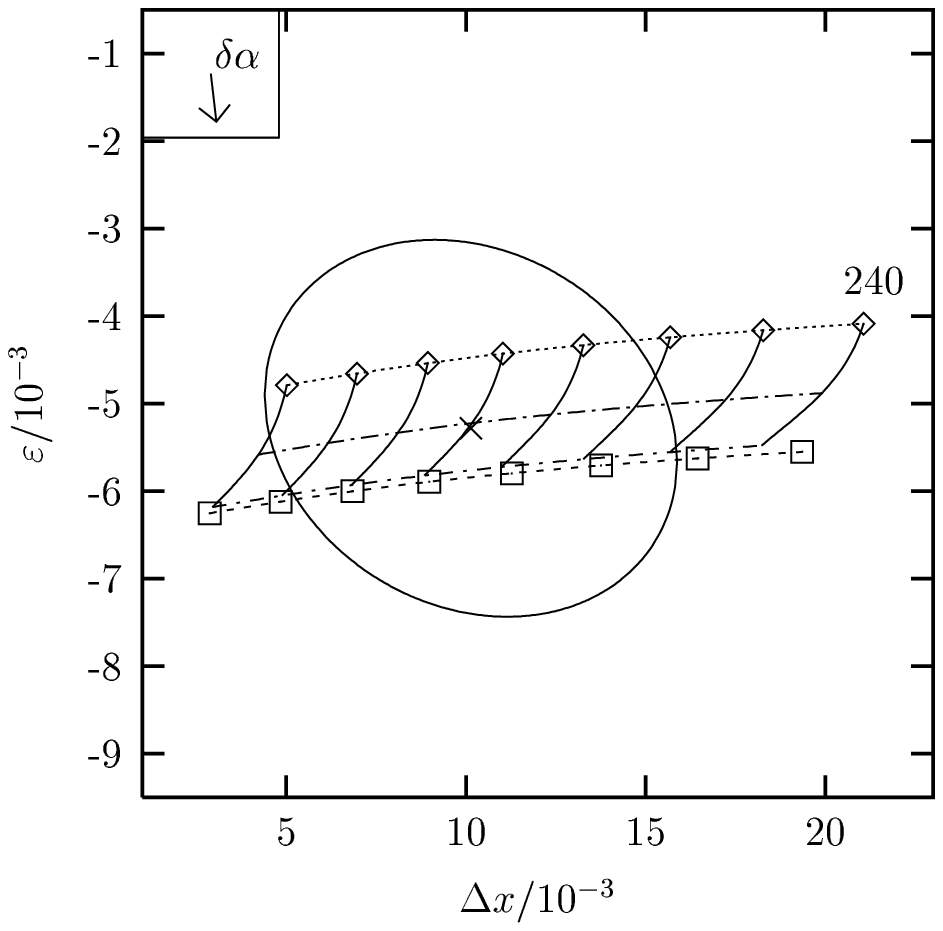}}
\put( 5.3, -6.0){\includegraphics{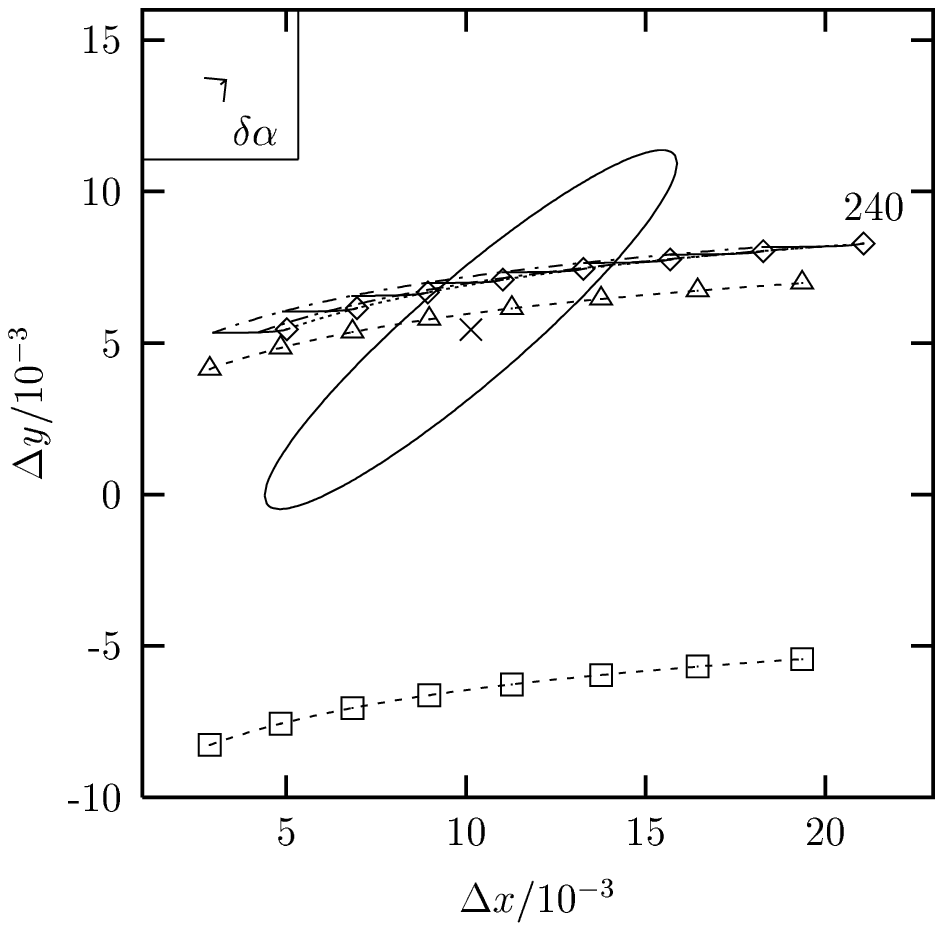}}
\put(-2.7,-14.4){\includegraphics{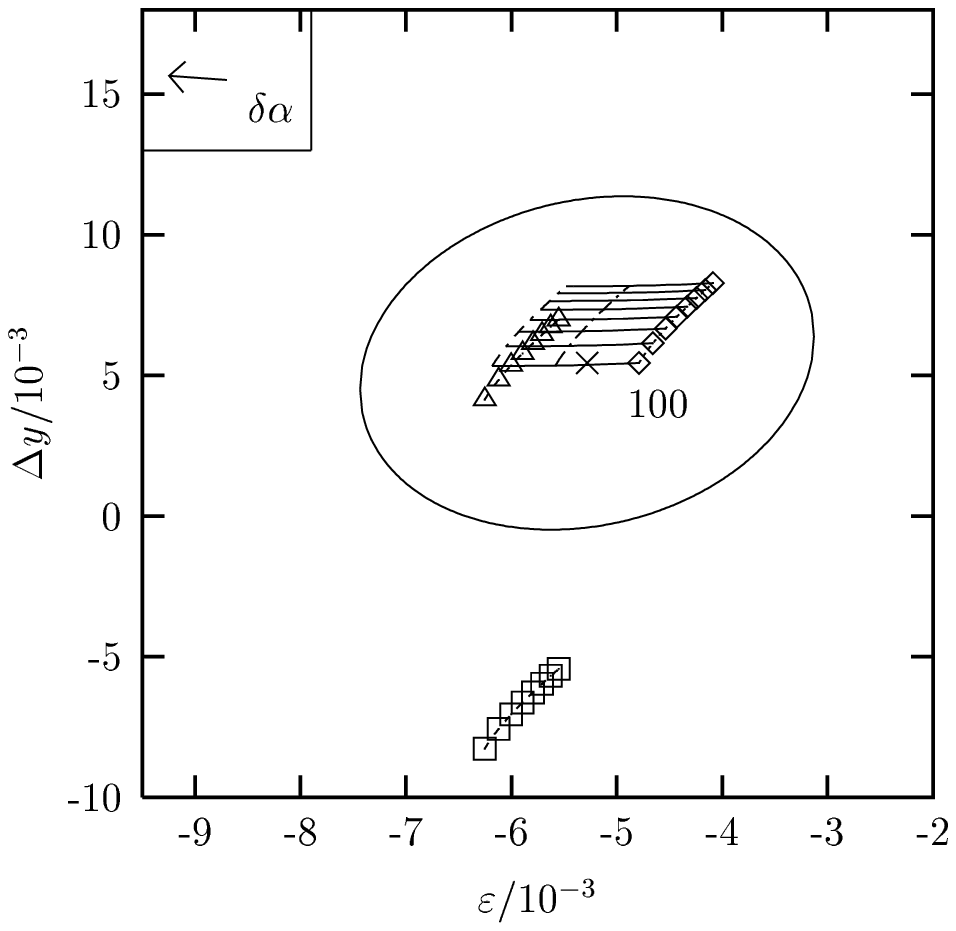}}
\end{picture}
\end{center}
\caption{
Comparison of the 83\% C.L.\ ($1.9 \sigma$) ellipses of the experimental data
with the full SM predictions and the pure
fermion-loop predictions in each plane of the three-dimensional
($\De x$, $\De y$, $\eps$)-space.
The full SM predictions are shown for Higgs-boson masses of
$100\GeV$ (dotted with diamonds), $300\GeV$ (long-dashed--dotted), $1\TeV$
(short-dashed--dotted) parametrized by the top-quark mass ranging from
$100$-$240\GeV$ in steps of $20\GeV$. The pure fermion-loop predictions
(short-dashed curves with squares) are shown
for the same top-quark masses. The additional dashed
curves with triangles correspond to $(\De y_{\fer} + \De y^{\SC}_{\bos})$
(see \refse{sec:dely}).
The arrow in the upper left corner of each figure
indicates how the empirical ellipse is shifted by the replacement
$\alpz^{-1}\to\alpz^{-1}+\de\alpz^{-1}$.}
\label{paramfig}
\efi

In \reffi{paramfig} the experimental results for the
parameters $\De x$, $\De y$, and $\eps$ are compared with the
theoretical predictions of the full SM and with the pure fermion-loop
predictions. In the theoretical predictions the leading two-loop
contributions of order $\ord(\alps\al t)$ and $\ord(\al^2t^2)$
have also been included (see \citere{zph2}).
The arrow in the upper
left corner of each figure indicates how the empirical 83\%
C.L.\ (i.e.\ $1.9 \sigma$) ellipse in the corresponding plane of
($\De x$, $\De y$, $\eps$)-space is shifted by the replacement
$\alpz^{-1}\to\alpz^{-1}+\de\alpz^{-1}$. In the theoretical
predictions for the effective parameters the error in $\alpz$
enters only in higher orders and is therefore negligible.

Figure \ref{paramfig} shows that the experimental
results are in excellent agreement with the full SM predictions
for top-quark masses which are compatible with the value of
$\Mt=180\pm 12\GeV$ \cite{mtexp} resulting from the
experimental detection at CDF and D\O. For the parameters $\De x$
and $\eps$ the data are consistently described by the pure
fermion-loop predictions alone, i.e.\ omission of the SM bosonic
corrections does not lead to a deviation from the experimental
results for these parameters. For the parameter $\De y$, on the
other hand, the pure fermion-loop prediction alone does not
yield a consistent description of the data, i.e.\ in this
parameter
additional corrections, such as the standard bosonic ones,
are required for an agreement between theory and experiment.
Since $\De y_{\fer}$ can reliably be calculated from the
experimentally well-known couplings of leptons and quarks,
this shows that the data have indeed become sensitive to
bosonic radiative corrections. These
significant
bosonic corrections are localized
in the single effective parameter $\De y$. It is
interesting to note that $\De y$ is totally insensitive to the
Higgs sector of the SM. It does not even show a logarithmic
dependence for a large Higgs-boson mass (see \citeres{zph3,gnlsm}).

The extra curves with triangles included in \reffis{paramfig}b and
\ref{paramfig}c
indicate the sum $(\De y_\fer+\De y_\bos^\SC)$, which besides
the fermion-loop contributions also includes the bosonic
corrections associated with the change in energy scale from
muon decay to W-boson decay, as will be discussed in
\refse{sec:dely}.

\section{\boldmath{Scale-change and isospin-breaking contributions to
the parameter $\De y$}}
\label{sec:dely}

As explained in the last section (see \refeq{y} and \refeq{eq:gw0}),
the effective parameter $\De y$, which incorporates the sizable bosonic
loop corrections, describes
both the change from 0 to $\MZ$ in the
energy scale and the transition from the charged-current to the
neutral-current
coupling. These two effects can separately be investigated by
introducing the charged-current
coupling $g_\PWpm(\MWpm^2)$ at the W-boson mass shell.
This coupling may be
defined via the leptonic width $\Ga^{\PW}_{\mathrm l}$ of the W~boson,
\beq
g^2_\PWpm(\MWpm^2) \equiv \frac{48 \pi}{\MWpm} \Ga^{\PW}_{\mathrm l}
\left(1 + c_0^2 \frac{3 \al}{4 \pi} \right)^{-1} .
\label{eq:Wlepwidth}
\eeq
In analogy to~\refeq{y} we relate $g_\PWpm(\MWpm^2)$ to
$g_\PWO (\MZ^2)$ by a parameter $\De y^{\IB}$,
\beq
\label{eq:delyib}
g^2_\PWpm(\MWpm^2) = y^{\IB} g^2_\PWO(\MZ^2)
= (1+\De y^{\IB})g^2_\PWO (\MZ^2) ,
\eeq
where the index ``IB'' refers to weak ``isospin-breaking''.
In \refeq{eq:Wlepwidth} we have introduced a factor
$(1 + c_0^2 3 \al/(4 \pi) )$ by convention. It is related to the
convention chosen in the treatment of the photonic corrections to the
leptonic $\PZ$-boson decay width $\Gl$ (see \refeq{obsleplow}).
The photonic contributions to $\Gl$ are pure QED corrections
giving rise to a
factor $(1 + 3 \al/(4 \pi))$ that is split off and not
included in $\De x$, $\De y$, and $\eps$.
In order to treat the photonic corrections on the same footing in both
the neutral and charged vector boson decay, one has to split off an
analogous correction factor also for the decay of the $\PW$~boson
(see \citere{zph3}).
The appearance of $c_0^2$ in the correction factor of
\refeq{eq:Wlepwidth}
is due to the rotation in isospin space relating the
physical field
$Z$ to the field $W^0$ entering the SU(2)
isotriplet. Numerically the correction term introduced in
\refeq{eq:Wlepwidth} amounts to $c_0^2 3\al/(4 \pi)=1.3\times 10^{-3}$.
Even though the convention chosen in \refeq{eq:Wlepwidth} is
well justified, it is worth noting that a different treatment of the
photonic corrections, e.g.~omission of the correction factor in
\refeq{eq:Wlepwidth},
would only lead to minor changes that do not influence our final conclusions.

The transition from the charged-current coupling at the scale of the
muon mass, $g_\PWpm(0)$, to the charged-current coupling obtained
from the decay of the W~boson into leptons, $g_\PWpm(\MWpm^2)$, can be
expressed by a parameter $\De y^{\SC}$,
\beq
g^2_\PWpm(0) = y^{\SC}g^2_\PWpm(\MWpm^2)
= (1+\De y^{\SC})g^2_\PWpm(\MWpm^2) ,
\label{eq:delysc}
\eeq
where the index ``SC'' means ``scale change''.
Inserting \refeq{eq:delyib} into \refeq{eq:delysc} and comparing with
\refeq{y}, one finds that
in linear approximation the parameter $\De y$ is
split into two additive contributions,
\beq
\label{eq:dely2}
\De y = \De y^{\SC} + \De y^{\IB} ,
\eeq
which furnish the transition from $g^2_\PWpm(0)$ to $g^2_\PWpm(\MWpm^2)$
and from $g^2_\PWpm(\MWpm^2)$ to $g^2_\PWO (\MZ^2)$, respectively.
Upon substituting \refeq{eq:gw0} and \refeq{eq:Wlepwidth} in
\refeq{eq:delysc}, one finds 
\beq
\label{eq:GaWySC}
\Delta y^{\SC} =
\frac{\MWpm^3\GF}{6\sqrt{2}\pi\Ga^{\PW}_{\mathrm l}} - 1
+c_0^2 \frac{3 \al}{4 \pi} ,
\eeq
which allows to determine $\Delta y^{\SC}$
(and consequently also $\Delta y^{\IB}$)
both experimentally and
theoretically. 

It should be noted at this point that the scale-change effect
discussed here does not correspond to an ordinary ``running'' of universal
(propagator-type) contributions, as
$g_\PWpm(0)$ and $g_\PWpm(\MWpm^2)$, being defined with reference to
muon decay and $\PW$-boson decay, respectively, are obviously
process-dependent quantities. Accordingly, the bosonic contributions to
$\De y^{\SC}$ (and also to $\De y$ and $\Delta y^{\IB}$) are
process-dependent. As these three parameters are
directly related to observables,
i.e.\ to complete S-matrix elements,
they are manifestly gauge-independent.

\btab
$$\begin{array}{|c||c|c|c|}
\hline
\Mt/\GeV & \De y_{\fer}/10^{-3} &
\De y^{\SC}_{\fer}/10^{-3} &
\De y^{\IB}_{\fer}/10^{-3} \\ \hline\hline
120 & -7.57 & -7.42 & -0.15 \\ \hline
180 & -6.27 & -7.79 & \phantom{-}1.52 \\ \hline
240 & -5.44 & -7.90 & \phantom{-}2.46 \\ \hline
\earr$$
$$\begin{array}{|c||c|c|c|}
\hline
\MH/\GeV & \De y_{\bos}/10^{-3} &
\De y^{\SC}_{\bos}/10^{-3} &
\De y^{\IB}_{\bos}/10^{-3} \\ \hline\hline
100 & 13.72 & 12.47 & 1.25 \\ \hline
300 & 13.62 & 12.42 & 1.20 \\ \hline
1000 & 13.61 & 12.41 & 1.20 \\ \hline
\earr$$
\caption{Fermionic and bosonic contributions to $\De y$, $\De y^{\SC}$,
and $\De y^{\IB}$ for different values of $\Mt$ and $\MH$.}
\label{ta:delY}
\etab

The analytical results of the SM predictions for $\De y^{\SC}$
and $\Delta y^{\IB}$ have been given in \citere{zph3}. As
it is the case for $\Delta y_{\bos}$, also $\Delta y^{\SC}_{\bos}$
and $\Delta y^{\IB}_{\bos}$ are insensitive to variations in
the Higgs-boson mass $\MH$.
In \refta{ta:delY} numerical results for the fermionic and bosonic
contributions to $\De y$, $\De y^{\SC}$, and $\De y^{\IB}$
are given for different values of $m_\Pt$ and $\MH$.
The table shows that both $\De y_{\fer}$ and $\De y_{\bos}$ are
dominated by the scale-change contributions
$\De y^{\SC}_{\fer}$ and $\De y^{\SC}_{\bos}$, respectively.
As these contributions enter with different signs, there are
strong cancellations in $\De y^\SC$ and $\De y$.

\begin{figure}
\begin{center}
\begin{picture}(16,12)
\put(-2.56,-14.2){\includegraphics{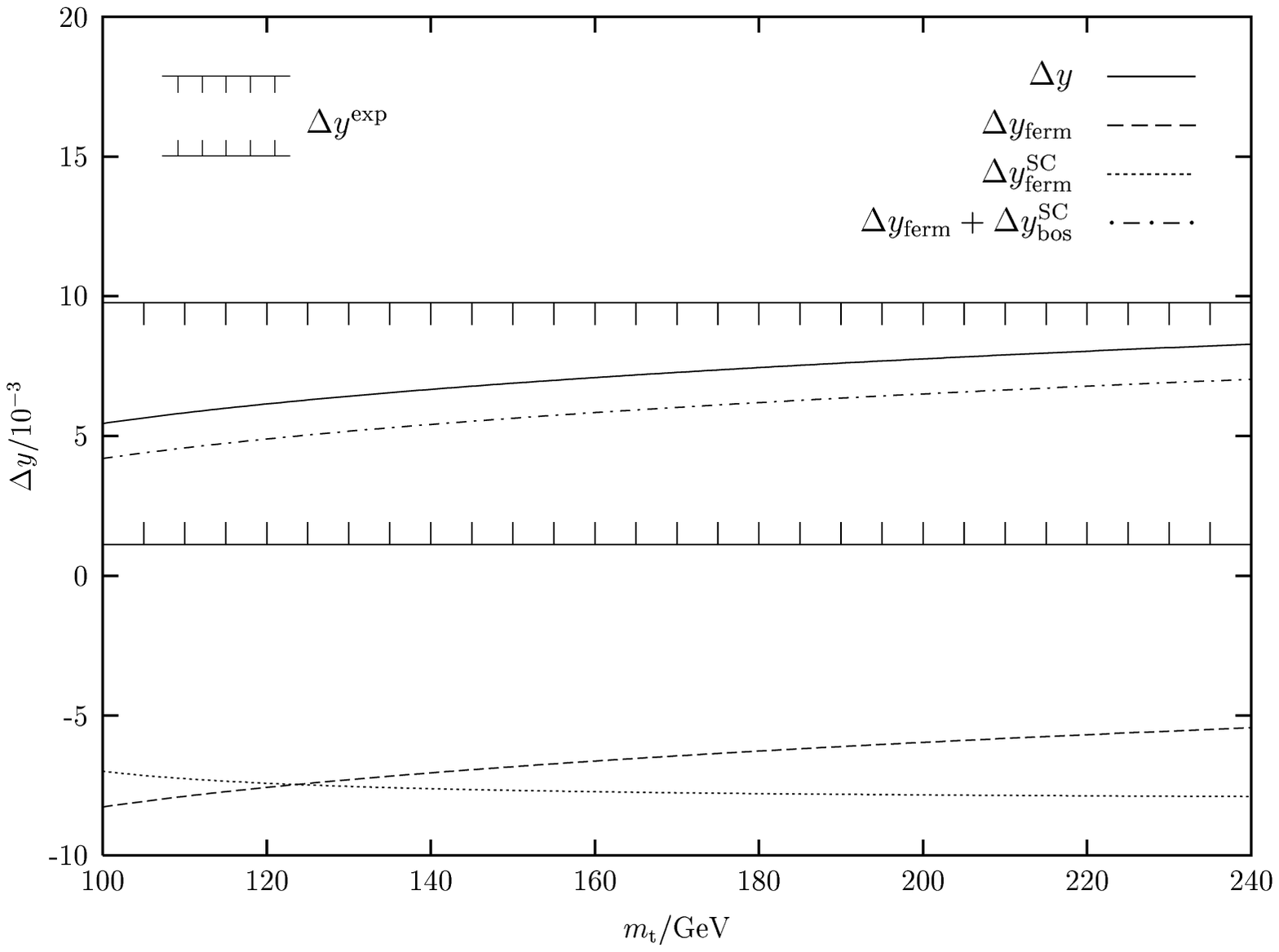}}
\end{picture}
\end{center}
\caption{The one-loop SM predictions for $\De y$, $\De
y_{\fer}$, $\De y^{\SC}_{\fer}$, and $(\De y_{\mathrm
ferm} + \De y^{\SC}_{\bos})$ as a function of $m_\Pt$.
The difference between the curves for $\De y$ and $(\De y_{\fer} +
\De y^{\SC}_{\bos})$ corresponds to the small contribution of $\De
y^{\IB}_{\bos}$.
The experimental value of $\Delta y$,
$\Delta y^{\mathrm exp} = (5.4 \pm 4.3) \times 10^{-3}$, is indicated
by the error band.}
\label{fig:delysc}
\efi

In \reffi{fig:delysc} we have plotted
the SM one-loop result for
$\De y$, $\De y_{\fer}$, $\De y^{\SC}_{\fer}$,
and $(\De y_{\fer} + \De y^{\SC}_{\bos})$ as a function
of $m_\Pt$ (using $\MH = 300 \GeV$).
The error band in \reffi{fig:delysc} indicates the
experimental value of $\Delta y$
given in \refeq{expxyeps}.
Fig.~\ref{fig:delysc} first of all displays
the above-mentioned fact that
the pure fermion-loop contribution
$\Delta y_\fer$ is not sufficient to
achieve agreement with $\Delta y^{\mathrm exp}$.
The contribution of
$\De y^{\SC}_{\fer}$ is approximately constant while
$\De y_{\fer} =
\De y^{\SC}_{\fer} + \De y^{\IB}_{\fer}$
shows a $\log\left(m_\Pt\right)$-dependence
which enters through $\De y^{\IB}_{\fer}$ (see \citere{zph3}).
In contrast to $\De y_{\fer}$, the complete one-loop result,
$\Delta y = \De y_{\fer} + \De y_{\bos}$, is in
perfect accordance with the data.
Figure \ref{fig:delysc} furthermore visualizes that $\De y^{\SC}_{\bos}$
constitutes by far the dominant part of the
bosonic contributions to $\Delta y$.
Combining the complete fermionic contribution $\De y_{\fer}$ with
the bosonic scale-change contribution
$\De y^{\SC}_{\bos}$ leads to a consistent description of the
current precision data, while the contribution of
$\De y^{\IB}_{\bos}$
does not give rise to a significant effect.

As a consequence of these
results on $\De y$ and of the results of \refse{sec:params}
on $\De x$ and $\eps$
we find that the effective parameters
$\De x$, $\De y$, $\eps$
are well approximated by
\beq
\De x\approx\De x_\fer, \qquad
\De y\approx\De y_\fer+\De y_\bos^\SC, \qquad
\eps\approx\eps_\fer, \quad
\label{eq:xyeapp}
\eeq
i.e.\ supplementing the fermion-loop approximation by the bosonic
scale-change contribution $\De y_\bos^\SC$ to $\De y$ leads to results
that deviate from the complete one-loop prediction for the effective
parameters by less than the experimental errors.

It is worth demonstrating this fact not only for the effective
parameters but also explicitly for the observables
$\bar\sw^2$, $\Gl$, $\MWpm/\MZ$. This is done in
\reffi{sm3d}, where the 68\% C.L.~($1.9 \sigma$)
volume defined by the
most recent 1995 data~\refeq{scdata} in the three-dimensional
($\MWpm/\MZ$, $\bar\sw^2$, $\Gl$)-space is shown together with the full
SM prediction, the pure fermion-loop prediction and the fermion-loop
prediction supplemented by the
bosonic contribution $\De y_\bos^\SC$.
For completeness the $\alpz$-Born
approximation is also indicated in the plots.
The error bars shown for the $\al(\MZ^2)$-Born approximation also apply
to all other theoretical predictions.
They originate from the errors of
the input parameters and are dominated by the error of $\alpz$ given
in \refeq{alpz1}.
The projections of the 68\% C.L.~volume onto the planes of the
three-dimensional ($\MWpm/\MZ$, $\bar\sw^2$, $\Gl$)-space correspond to
the 83\% C.L.\ ellipse in each plane,
while the projections onto the individual axes correspond to the 94\%
C.L.\ there.

\newcounter{subfigure}
\setcounter{subfigure}{1}
\makeatletter
\def\thesubfigure{\@alph\c@subfigure}
\def\fnum@figure{\figurename~\thefigure\thesubfigure}
\newcommand{\sfcount}{\addtocounter{subfigure}{1}\addtocounter{figure}{-1}}
\makeatother
\begin{figure}
\begin{center}
\begin{picture}(16,13)
\put(0.7,6.5){$\bar\sw^2$}
\put(2,0.8){$\MWpm/\MZ$}
\put(11.4,0.6){$\Gl/\MeV$}
\put(-1.4,-6.0){\includegraphics{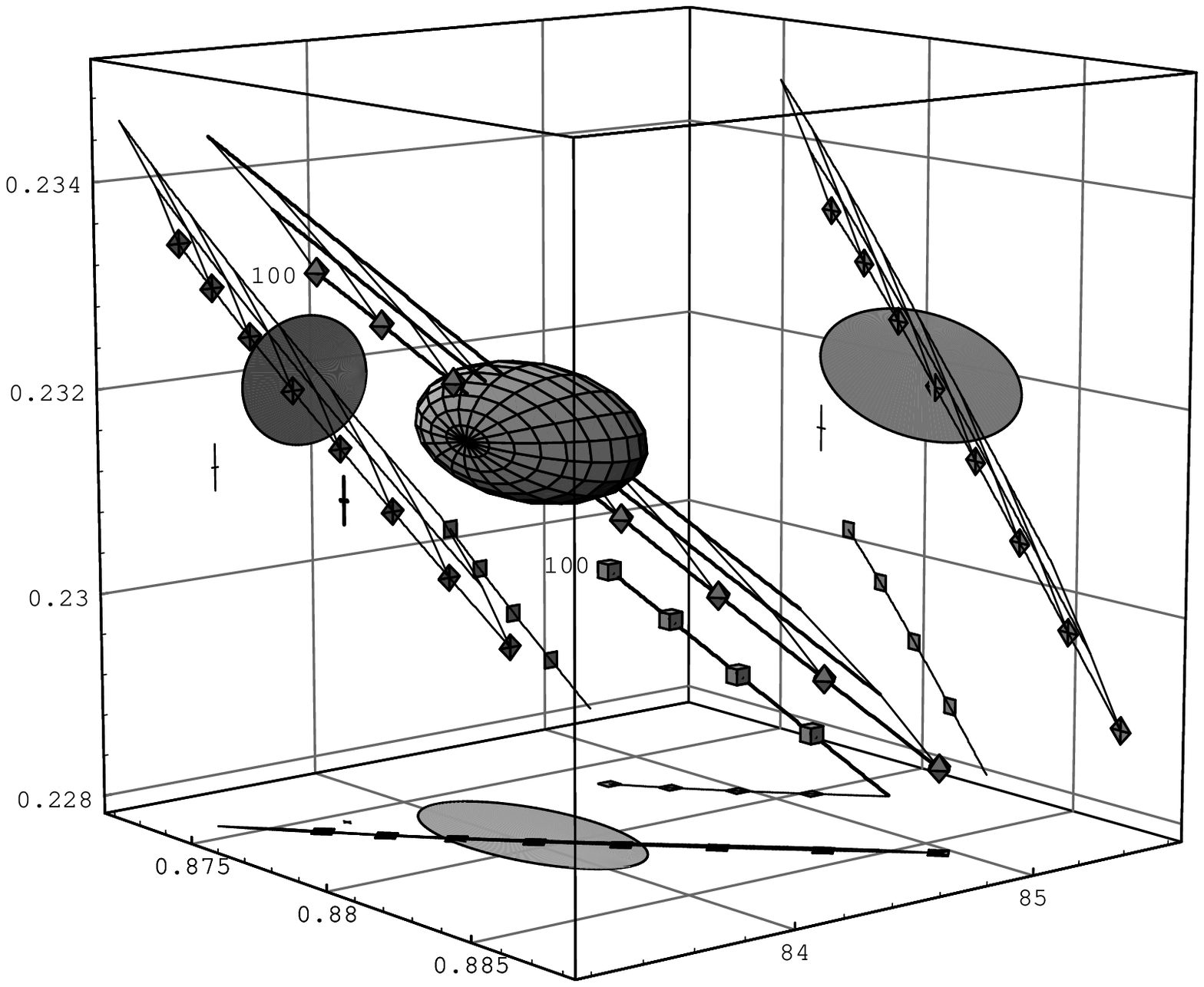}}
\end{picture}
\end{center}
\caption[]{Three-dimensional plot of the 68\% C.L.~($1.9 \sigma$) ellipsoid
of the experimental data in ($\MWpm/\MZ$, $\bar\sw^2$, $\Gl$)-space
and comparison with the full SM prediction (connected lines)
and the pure fermion-loop prediction (single line with cubes).
The full SM prediction is shown for Higgs-boson masses of $\MH = 100
\GeV$ (line with diamonds), $300\GeV$, and
$1\TeV$ parametrized by $\Mt$
ranging from 100--240$\GeV$ in steps of $20\GeV$. In the pure
fermion-loop prediction the cubes also indicate steps in $\Mt$ of
$20\GeV$ starting with $\Mt = 100$ $\GeV$.
The cross outside the ellipsoid indicates the $\al(\MZ^2)$-Born
approximation with the corresponding error bars,
which also apply to all other theoretical predictions.}
\sfcount
\label{sm3d}
\efi
\begin{figure}
\begin{center}
\begin{picture}(16,13)
\put(0.7,6.5){$\bar\sw^2$}
\put(2,0.8){$\MWpm/\MZ$}
\put(11.4,0.6){$\Gl/\MeV$}
\put(-1.4,-6.0){\includegraphics{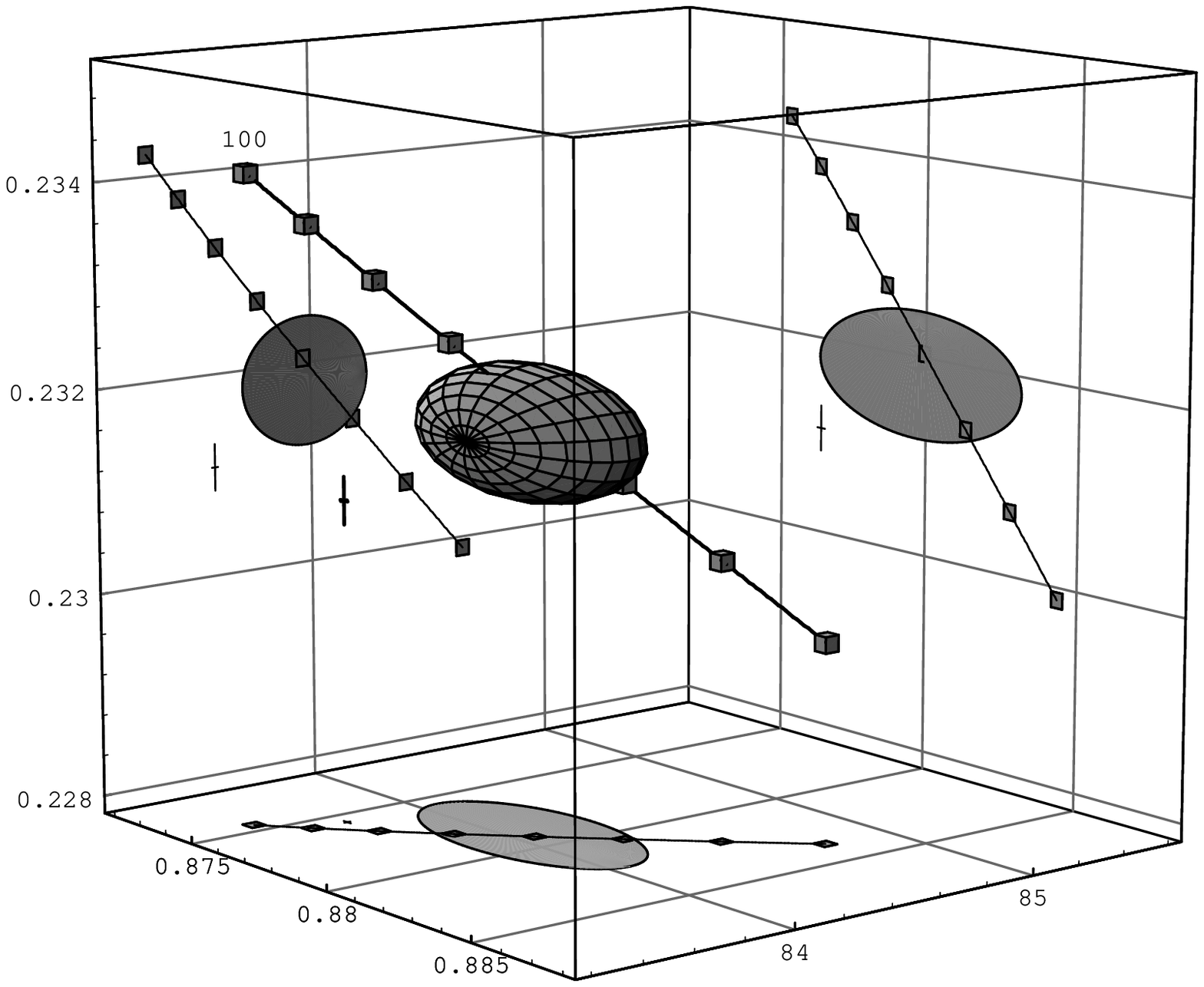}}
\end{picture}
\end{center}
\caption[]{Three-dimensional plot of the 68\% C.L.~($1.9 \sigma$) ellipsoid
of the experimental data in ($\MWpm/\MZ$, $\bar\sw^2$, $\Gl$)-space
and comparison with the theoretical prediction obtained by combining
the fermion-loop contribution with the
bosonic correction $\De y_\bos^\SC$ related to the scale change from
$\GF$ to $\Ga^{\PW}_\Pl$.
The theoretical prediction
is parametrized by $\Mt$ ranging from 100--240$\GeV$ in
steps of $20\GeV$.}
\efi
\makeatletter
\def\fnum@figure{\figurename~\thefigure}
\makeatother
In \reffi{sm3d}a the single line shows the theoretical prediction
for values of $\Mt$ varying from $\Mt=100-240\GeV$ taking
into account only fermion-loop corrections.
The three connected lines show the full SM predictions.
They correspond to $\MH = 100, 300, 1000 \GeV$, respectively,
and $\Mt$ is again varied from $\Mt=100-240\GeV$.
In both theoretical predictions the leading two-loop
contributions of order $\ord(\alps\al t)$ and $\ord(\al^2t^2)$
have been included (see \citere{zph2}).
The full SM prediction is in agreement with the data
for the empirical value of the top-quark mass,
$\Mt = 180 \pm 12 \GeV$~\cite{mtexp}.
The pure fermion-loop prediction, on the other hand, differs from the
data by several standard deviations, which clearly illustrates
the sensitivity of the data to SM bosonic loop effects.
For $\Mt = 180 \GeV$ the pure fermion-loop predictions
at one-loop order read
\beqar
\bar s^2_{{\ss\PW}, {\fer}} &=& 0.22747 \mp 0.00023, \nn\\
\left(\frac{\MWpm}{\MZ}\right)_{\fer} &=& 0.88358 \pm 0.00013, \nn\\
\Ga_{\Pl, {\fer}} &=& 85.299 \pm 0.012\MeV,
\label{eq:ferpred}
\eeqar
which deviate from the experimental values by $-13\si$,
$1.9\si$ and $9.8\sigma$,
respectively.
As mentioned above, all uncertainties of theoretical predictions
are dominated by the error of $\alpz$ given in \refeq{alpz1}.
The $\alpz$-Born approximations for the observables read
\beqar
s_0^2     &=& 0.23112 \mp 0.00023, \nn\\
c_0       &=& 0.87686 \pm 0.00013,   \nn\\
\Gl^{(0)} &=& 83.563 \pm 0.012\MeV,
\label{eq:alpzborn}
\eeqar
corresponding to a deviation of $-2.2\si$,
$-1.9\si$ and $-2.6\si$, respectively, from the experimental data.
The fact that the values in \refeq{eq:alpzborn} are closer to the empirical
data and the full SM predictions than the fermion-loop prediction
\refeq{eq:ferpred} is a consequence of the cancellation
between fermionic and bosonic contributions in the single parameter
$\De y$.

Figure \ref{sm3d}b shows the
theoretical predictions obtained by combining the pure
fermion-loop prediction with the
bosonic contribution
$\De y_\bos^\SC$ according to \refeq{eq:xyeapp}. As expected
from \reffi{fig:delysc},
adding only the bosonic scale-change
contribution $\De y_\bos^\SC$ to the fermion-loop contribution
is sufficient to obtain a theoretical prediction that is in
agreement with the data. It is very close to the full SM
prediction, i.e.\ the difference between these predictions is
below experimental resolution.

We therefore conclude that the bosonic corrections required
for consistency between the precision data for the observables
$\bar\sw^2$, $\Gl$, $\MWpm/\MZ$ and the theoretical predictions
in terms of the input parameters $\al(\MZ^2)$, $\MZ$ and $\GF$
are just those contained in the parameter $\De y_\bos^\SC$.
As $\De y^\SC$ describes the transition between $g^2_\PWpm(0)$,
i.e.\ $\GF$, and $g^2_\PWpm(\MWpm^2)$, i.e.\ $\Ga^{\PW}_{\Pl}$,
the bosonic corrections required by the
precision data can be identified as an effect of changing the
energy scale from the low-energy process muon decay to the
energy scale of W-boson decay.
All other bosonic effects, in particular the $\log(\MH)$-dependent
vacuum-polarization contributions contained in $\De x_\bos$ and
$\eps_\bos$, are below experimental resolution for Higgs-boson masses in
the perturbative regime, i.e.\ below $\sim 1\TeV$.

The sensitivity of the data on variations in the Higgs-boson mass can be
inspected in \reffi{sm3d}a.
If $\Mt$ is fixed, the intersection of the three-dimensional
68\% C.L.\ ($1.9 \sigma$) volume with
the lines representing the full SM prediction shows a certain
sensitivity to the Higgs-boson mass. It can also be seen,
however, that in the direction in three-dimensional space in
which the $\MH$-dependence
(for fixed $\Mt$) is sizable also the uncertainty in
the theoretical predictions due to the error in $\al(\MZ^2)$ is large.

\section{\boldmath{Radiative corrections in the
$\Ga^{\PW}_{\mathrm l}$-scheme}}
\label{sec:gawscheme}

After having identified the source of the important
bosonic corrections in the analysis of the precision data
as a scale-change effect related to the use
of the low-energy input parameter $\GF$, it is evident that
these large bosonic corrections could be avoided by expressing
the theoretical predictions for the observables $\bar\sw^2$,
$\Gl$, $\MWpm/\MZ$ in terms of input parameters being defined at
the scale of the vector-boson masses.
This can be achieved by using the $\PW$-boson
width $\Ga^{\PW}_{\mathrm l}$ instead of the Fermi constant $\GF$
as an input parameter.

In the language of the effective Lagrangian ${\cal L}_{\mathrm C}$ given
in \refeq{lc} the use of the input quantity $\Ga^{\PW}_{\mathrm l}$ instead
of $\GF$ means that the charged-current coupling in \refeq{lc}
is identified with $g_\PWpm(\MWpm^2)$ defined via the $\PW$-boson width
$\Ga^{\PW}_{\mathrm l}$ (see \refeq{eq:Wlepwidth}) rather than with
$g_\PWpm(0)$ defined via muon decay (see \refeq{eq:gw0}).
With this identification
the transition between $g_\PWpm(0)$ and $g_\PWpm(\MWpm^2)$, and
accordingly the contribution of $\De y^\SC$, does not occur. The
radiative corrections to the observables
$\swbar^2$, $\Gl$, and $\MWpm/\MZ$ are completely contained in the
parameters $\De x$, $\De y^{\IB}$, and $\eps$, in which SM
corrections beyond fermion loops do not give rise to significant
contributions.

In this ``$\Ga^\PW_\Pl$-scheme'' the lowest-order values
$\hat s_0^2$, $\hat c_0$, and $\hat \Gl^{(0)}$ of the
observables are given in terms
of the input quantities $\al (\MZ^2)$, $\MZ$, and $\Ga^\PW_\Pl$
as
\beq
\frac{\hat s_0^2}{\hat c_0} \equiv \frac{\al (\MZ^2) \MZ}{12
\Ga^{\PW}_{\mathrm l}} \left(1 + c_0^2 \frac{3 \al}{4 \pi}\right),
\qquad \hat c_0^2 \equiv (1 - \hat s_0^2),
\eeq
and
\beq
\hat\Gl^{(0)} = \frac{\alpz\MZ}{48 \hat s_0^2 \hat c_0^2}
\left[1+(1-4 \hat
s_0^2)^2\right]\left(1+\frac{3\alpha}{4\pi}\right) .
\eeq
The linearized relations between the observables and the effective
parameters $\De x$, $\De y^\IB$, and $\eps$ read
\beqar
\bar\sw^2 &=& \hat s_0^2\left[1+
\frac{\hat c_0^2}{2- \hat s_0^2} \De x +
\frac{2 \hat c_0^2}{2- \hat s_0^2} \De y^{\IB} +
\frac{3\hat s_0^2 - 2}{2- \hat s_0^2} \eps
+ (\hat c_0^2 - \hat s_0^2) \de \right], \nn\\
\frac{\MWpm}{\MZ} &=& \hat c_0\left[1+
\frac{\hat c_0^2}{2- \hat s_0^2} \De x -
\frac{\hat s_0^2}{2- \hat s_0^2} \De y^{\IB} +
\frac{2 \hat s_0^2}{2- \hat s_0^2} \eps
\right], \nn\\
\Gl &=& \hat \Gl^{(0)}\biggl[1 -
\frac{2}{(2- \hat s_0^2) \left[1 + (1 - 4 \hat s_0^2)^2 \right]}
\left((1 - 2 \hat s_0^2 - 4 \hat s_0^4) (\De x + 2 \De y^{\mathrm
IB})
\right. \nn\\
&& \left. {} \qquad \qquad
- 2 \hat s_0^2 (1 - 10 \hat s_0^2) \eps - 8 \hat s_0^4 (2- \hat
s_0^2) \de
\right) \biggr] .
\label{eq:analWlin}
\eeqar

The relations \refeq{eq:analWlin} could in principle be used for
a data analysis of the observables $\bar\sw^2$, $\Gl$, and
$\MWpm/\MZ$ in the $\Ga^{\PW}_{\Pl}$-scheme, i.e.~with
$\al(\MZ^2)$, $\MZ$, and $\Ga^{\PW}_{\Pl}$ as experimental input
quantities. Assuming (hypothetically) the same experimental
accuracy as in the ``$\GF$-scheme'' (input parameters
$\al(\MZ^2)$, $\MZ$, and $\GF$) and an experimental value of
$\Ga^{\PW}_{\Pl}$ being in agreement with the SM prediction,
a consistent description of the
data in the $\Ga^{\PW}_{\Pl}$-scheme would be possible by solely
including the pure fermion-loop predictions in the effective
parameters.

At present a data analysis using the $\Ga^{\PW}_{\Pl}$-scheme
would of course not be sensible owing to the large experimental
error in the determination of the $\PW$-boson width. {}From
\citere{PDB} we have $\Ga^{\PW,\mathrm exp}_{\mathrm T} = (2.08 \pm
0.07) \GeV$ for the total decay width of the W-boson and $(10.7
\pm 0.5) \%$ for the leptonic branching ratio. Adding the errors
quadratically yields $\Ga^{\PW,\mathrm exp}_{\mathrm l} = (223 \pm 13)
\MeV$ showing that the experimental error in $\Ga^{\PW,\mathrm
exp}_{\mathrm l}$ at present is
more than one order of magnitude larger
than the error in the leptonic \PZ-boson width (see
\refeq{scdata}) and obviously much larger than the one in $\GF$
(see \refeq{eq:GF}).

Even though a precise experimental input value for $\Ga^\PW_\Pl$
is not available, it is nevertheless instructive to simulate the
analysis in the $\Ga^\PW_\Pl$-scheme by using the theoretical SM
value for $\Ga^\PW_\Pl$ as hypothetical input parameter for
evaluating \refeq{eq:analWlin}. For the choice of $\Mt = 180
\GeV$ and $\MH = 300 \GeV$ one obtains
$\Ga^{\PW}_{\Pl} = 226.3 \MeV$
as theoretical value of $\Ga^\PW_\Pl$ in the SM.
One should note that our procedure here is technically analogous
to commonly used parametrizations of radiative corrections
where, for instance in
the on-shell scheme (see e.g.~\citere{adhabil}), the
corrections are expressed in terms of the $\PW$-boson mass
$\MWpm$, while in an actual evaluation $\MWpm$ is substituted by
its theoretical SM value in terms of $\alpz$, $\MZ$, and $G_\mu$.

\btab
\renewcommand{\arraystretch}{1.7}
\arraycolsep 5pt
$$\begin{array}{|c||c|c||c|c||c|}
\hline
& \multicolumn{2}{|c||}{\GF{\mathrm -scheme}} &
  \multicolumn{2}{c||}{\Ga^{\PW}_{\mathrm l}{\mathrm -scheme}} &
\\ \cline{2-5}
& {\mathrm ferm.~corr.} & {\mathrm bos.~corr.} &
  {\mathrm ferm.~corr.} & {\mathrm bos.~corr.} &
  {\mathrm exp.~error} \\ \hline\hline
\frac{\De\bar\sw^2}{\bar\sw^2}/ 10^{-3} &
-15.8 & \phantom{-}16.3 & \phantom{-}11.0 & \phantom{-}1.3 &
1.5 \\ \hline
\frac{\De\MWpm}{\MWpm} /10^{-3} &
\phantom{-1}7.7 & ~-1.6 & \phantom{-1}3.7 &
\phantom{-}0.6 & 2.0 \\ \hline
\frac{\De \Gl}{\Gl} / 10^{-3} &
\phantom{-}20.8 & -14.3 & ~-1.8 & -1.7 & 1.7 \\ \hline
\earr$$
\caption{
Relative size of the SM fermionic and bosonic one-loop corrections
to the observables $\bar\sw^2$, $\MWpm$, and $\Gl$ in the $\GF$-scheme
(input parameters $\GF$, $\MZ$, and $\al (\MZ^2)$)
and in the simulated $\Ga^{\PW}_{\mathrm l}$-scheme
(input parameters $\Ga^{\PW}_{\Pl} = 226.3 \MeV$, $\MZ$,
and $\al (\MZ^2)$)
for  $\Mt = 180 \GeV$ and $\MH = 300 \GeV$.
The relative experimental error of the observables is also indicated.}
\label{tab:analW}
\etab

In order to illustrate the fact that the replacement of the input
quantity $\GF$ by $\Ga^{\PW}_{\mathrm l}$ indeed strongly affects the
relative size of the fermionic and bosonic contributions
entering each observable,
we have given in \refta{tab:analW} the
relative values of the SM one-loop fermionic and bosonic
corrections to the observables $\bar\sw^2$, $\MWpm$, and $\Gl$
in the $\GF$-scheme and in the simulated $\Ga^{\PW}_{\mathrm
l}$-scheme based on the input value $\Ga^{\PW}_{\Pl} = 226.3
\MeV$. The size of the radiative corrections in the two schemes
is compared with the relative experimental error of the observables
(see \refeq{scdata}).
Table \ref{tab:analW} shows that in the $\GF$-scheme the bosonic
corrections to $\bar\sw^2$ and $\Gl$ are quite sizable and considerably
larger than the experimental error. In the (simulated)
$\Ga^{\PW}_{\mathrm l}$-scheme,
on the other hand, these corrections are smaller by
an order of magnitude
and have about the same size as the experimental
error. The bosonic contributions to $\MWpm$ are smaller than the
experimental error in both schemes.
It can furthermore be seen in \refta{tab:analW} that the
cancellation between fermionic and bosonic corrections related
to the scale change is not present in the $\Ga^{\PW}_{\Pl}$-scheme.

The explicit values for the pure fermion-loop predictions of the
observables
at one loop
in the simulated $\Ga^{\PW}_{\Pl}$-scheme read
\beq
\bar s^2_{{\ss \PW}, {\fer}} = 0.23154, \qquad
\left(\frac{\MWpm}{\MZ}\right)_{\fer} = 0.8813, \qquad
\Ga_{\Pl, {\fer}} = 84.06 \MeV. \quad
\eeq
Comparison with the experimental values of the observables given
in \refeq{scdata} shows that there are indeed no significant
deviations between the pure fermion-loop predictions in the
simulated $\Ga^{\PW}_{\Pl}$-scheme and the data,
i.e.\ they agree within one standard deviation. This has to be
contrasted to the situation in the $\GF$-scheme, where the pure
fermion-loop predictions differ from the data by several
standard deviations (see \refeq{eq:ferpred} and
\reffi{sm3d}a).

In summary, we have demonstrated that after replacing
the low-energy quantity $\GF$ by the high-energy observable
$\Ga^{\PW}_{\Pl}$ in the theoretical predictions
for the observables $\bar\sw^2$, $\MWpm/\MZ$, and $\Gl$,
no corrections beyond fermion loops are required in order to
consistently describe the data.
Although at present,
due to the large experimental error in $\Ga^{\PW}_{\Pl}$,
the so-defined $\Ga^{\PW}_{\Pl}$-scheme is
of no practical use
for analyzing the precision data,
from a theoretical point of view
it shows that the only bosonic corrections of
significant magnitude
are such that they can completely be absorbed by
the introduction of the quantity $\Ga^{\PW}_{\Pl}$.

\section{Conclusions}
\label{conc}

In this paper the nature of those
electroweak bosonic loop corrections that are significant
in the comparison between theory and present precision data
has been investigated.
The analysis has been based on the leptonic LEP1 observables $\Gl$
and $\bar\sw^2$, which are not influenced by the discrepancies noted
in certain hadronic decay modes of the Z~boson, and the $\PW$-boson
mass $\MWpm$. The experimental uncertainty in the hadronic sector enters
only via the input parameter $\alpz$. As further input parameters $\MZ$
and $\GF$ have been used.

For analyzing the structure of the bosonic corrections it is
particularly convenient to use the set of
effective parameters $\De x$, $\De y$ and $\eps$ being introduced on the
basis of an effective Lagrangian that quantifies different sources of
SU(2) violation. It has been shown that non-fermionic corrections are only
required in the single parameter $\De y$ which in turn is practically
independent of the Higgs sector of the theory.

By studying the bosonic contributions entering $\De y$ it has
furthermore been pointed out that
the bosonic corrections needed for an agreement between the SM
predictions and the current precision data
can be identified as an effect of the change in energy scale
from the low-energy process muon decay to the energy scale of
the LEP observables.
More precisely, the bosonic corrections
resolved by the precision experiments are just those
furnishing the transition from the low-energy parameter $\GF$ to the
leptonic width of the $\PW$~boson, $\Ga^{\PW}_{\mathrm l}$.
Upon introducing the high-energy quantity $\Ga^{\PW}_{\Pl}$
($\Ga^{\PW}_{\Pl}$-scheme) as input parameter instead of
$\GF$ ($\GF$-scheme), the relevant bosonic corrections
for the high-energy observables $\swbar^2$, $\Gl$, and $\MWpm/\MZ$
can be absorbed. This illustrates that the question whether bosonic
loop corrections are in fact needed in order to match theory and
experiment for a certain set of observables depends on the set of input
quantities, or in other words on the renormalization scheme,
one is using when evaluating the theoretical predictions for the
observables.
Since the usefulness of $\Ga^{\PW}_{\Pl}$ as experimental input parameter
is limited at present due to the large experimental error of the
W-boson width, we have demonstrated this fact by invoking the SM
theoretical value of $\Ga^{\PW}_{\mathrm l}$ as input. Indeed, no further
corrections beyond fermion loops are needed in this case in order to
achieve agreement with the data within one standard deviation.

\section*{Acknowledgement}
G.~W.\ thanks R.~R${{{\rm a}_{}}_{}}_{\hskip -0.18cm\varsigma}$czka
and the organizing committee of the symposium for invitation,
their warm hospitality and for providing a very pleasant atmosphere
during the workshop.

\end{document}